\documentclass[prl,twocolumn,showpacs]{revtex4}
\usepackage{graphicx}
\begin{document}
\title{Metallic ground state and glassy transport in single crystalline URh$_2$Ge$_2$: Enhancement of disorder effects in a strongly correlated electron system}
\author{S. S\"ullow$^1$, I. Maksimov$^1$, A. Otop$^1$, F.J. Litterst$^1$, A. Perucchi$^2$, L. Degiorgi$^2$, J.A. Mydosh$^{3,4}$}
\address{$^1$Institut f\"{u}r Metallphysik und Nukleare Festk\"{o}rperphysik, TU Braunschweig, 38106 Braunschweig, Germany\\
$^2$Laboratorium f\"{u}r Festk\"{o}rperphysik, ETH Z\"urich, Switzerland\\
$^3$Kamerlingh Onnes Laboratory, Leiden University, 2300 RA Leiden, The Netherlands\\
$^4$Max Planck Institute for Chemical Physics of Solids, 01187 Dresden, Germany}
\date{\today}

\begin{abstract}
We present a detailed study of the electronic transport properties on a single crystalline specimen of the moderately disordered heavy fermion system URh$_2$Ge$_2$. For this material, we find glassy electronic transport in a single crystalline compound. We derive the temperature dependence of the electrical conductivity and establish metallicity by means of optical conductivity and Hall effect measurements. The overall behavior of the electronic transport properties closely resembles that of metallic glasses, with at low temperatures an additional minor spin disorder contribution. We argue that this glassy electronic behavior in a crystalline compound reflects the enhancement of disorder effects as consequence of strong electronic correlations.
\end{abstract}


\maketitle

The interplay of disorder and electronic correlations represents a central issue of the physics of strongly correlated electron systems (SCES). The properties of SCES are extraordinarily sensitive to small levels of crystallographic disorder. This is attributed to the electronic correlations, which enhance the effect of the disorder~\cite{wysokinski}. Correspondingly, disorder has been proposed to greatly modify or control the physical properties of SCES in very many cases~\cite{maple,capogna}. A thorough understanding of this issue, however, be it from the material physics point of view~\cite{millis} or concerning the process of localization in correlated electron systems~\cite{belitz,dobro}, is lacking.

Heavy--fermion metals represent an archetypical class of intensively studied SCES~\cite{stewart}. For heavy fermions, disorder effects have been found to affect the behavior close to magnetic instabilities, like in Ce--based compounds~\cite{rosch} or in U--intermetallics as UCu$_4$Pd or (U,Th)Pd$_2$Al$_3$~\cite{miranda1}. Especially, the electronic transport properties of these U heavy fermions have been interpreted in terms of Non--Fermi--Liquid (NFL) behavior. Rather than exhibiting a generic heavy fermion metallic resistivity, with a $ln(T)$ dependence at high temperatures $T$ and a crossover below $T_{coh}$ into a coherent state with a metallic resistivity $\rho$\,=\,$\rho_0$\,+\,$A$\,$T^2$, for these materials a resistivity deviating from Fermi liquid predictions is found, with $\rho$\,=\,$\rho_0 \left[1-a\left(T/T^* \right)^n\right]$, $n$\,$\sim$\,1\,--\,1.5~\cite{andraka}. Yet, the resistivities of these and various related materials are highly unusual in other aspects as well, as they remain -- for metallic systems -- large (many hundred $\mu \Omega$cm), and exhibit a negative temperature coefficient of the resistivy (TCR), $d \rho /d T$\,$<$\,0, down to lowest temperatures~\cite{shlyk,sullow1,li}.

Metallurgically, these U compounds with a negative TCR are characterized by ''moderate'' disorder, that is crystallographic randomness on the atomic level of $\sim$\,10 vol.\,\%. Under such circumstances, the question about the origin of the anomalous transport properties arises: Is the negative TCR predominantly the result of quantum spin fluctuations and hence an indication for a NFL~\cite{miranda1}? Or is it the result of the combined influence of strong electronic correlations and disorder--induced localization, which ought to be treated in the framework of Anderson vs. Mott--Hubbard localization~\cite{lee,miranda2}. In order to address such issues we have performed a detailed study of the electronic transport properties of a U heavy fermion compound with moderate and known disorder, namely URh$_2$Ge$_2$~\cite{sullow2}.

Previously, we have established this material as the first 3D random--bond, heavy--fermion Ising--like spin glass~\cite{sullow2}. Further, we have determined type and level of the crystallographic disorder. The system crystallizes in the tetragonal CaBe$_2$Ge$_2$ lattice ({\it P4/nmm}). Disorder is present on the non--magnetic ligand sites in form of moderate bond length disorder, which likely is the result of $\sim$\,5\,--\,10\,\% Rh/Ge random site exchange~\cite{booth}. In contrast, the U ions occupy translationally invariant positions on an ordered tetragonal sub--lattice. The bond length disorder, in as--grown single crystals, generates the spin glass ground state below $T_f$\,$\simeq$\,9\,K.

The electronic transport even in single crystalline URh$_2$Ge$_2$ follows that of an archetypical moderately disordered U heavy fermion compound~\cite{sullow1}. The absolute resistivity $\rho$ along $a$ and $c$ axes ranges from 200 to 800\,$\mu \Omega$cm, with large sample--to--sample variations, while the TCR is negative up to well above 300\,K. Below $\sim$\,10\,K, in close resemblance to UCu$_4$Pd, the resistivity exhibits a NFL--like temperature dependence $\rho$\,=\,$\rho_0 \left[1-a T \right]$. While in UCu$_4$Pd the negative TCR has been interpreted within the framework of the disordered Kondo/Griffiths phase scenario, for various reasons~\cite{sullow1} in the spin glass URh$_2$Ge$_2$ this does not reflect a NFL ground state in the spirit of the Refs.~\cite{maple,miranda1}. First, a single ion Kondo--like behavior of the resistivity $\propto$\,$ln{T}$ is not observed. Secondly, the $\rho$ values in URh$_2$Ge$_2$ by far exceed those obtained from the Ioffe--Regel criterion, indicating substantial electronic localization. Thirdly, while the theoretical models in Ref.~\cite{miranda1} assume a distribution of Kondo temperatures $T_K$ from 0 to some finite value, due to the finite freezing transition temperature $T_f$ in URh$_2$Ge$_2$, such a wide $T_K$ distribution is unlikely. Finally, the anisotropic response of the resistivity to an annealing treatment, yielding a heavy fermion metallic behavior along the $a$ axis and a negative TCR along the $c$ axis, can not be reconciled with single ion Kondo scattering models or their extensions as put forth in Ref.~\cite{miranda1}. Thus, the mechanism which actually controls the electronic transport has not been resolved.

In order to elucidate the physical mechanisms behind this highly unusual behavior, we have performed a thorough study of the electronic transport properties of single crystalline spin glass URh$_2$Ge$_2$. In this Letter, we will present evidence for glassy electronic transport even in single crystalline URh$_2$Ge$_2$. We will extract a generic temperature dependence of the electrical conductivity $\sigma$ and establish the metallicity of the material by means of optical conductivity and Hall effect measurements. Based on its temperature and magnetic field dependence, we distinguish between two conductivity components. The overall behavior of $\sigma$ resembles that of metallic glasses, and correspondingly, we argue that it is primarily governed by disorder induced localization effects~\cite{lee}. In addition, at low $T$ there is a secondary contribution from magnetic (spin disorder) scattering.

\begin{figure}[!ht]
\begin{center}
\includegraphics[width=1\linewidth]{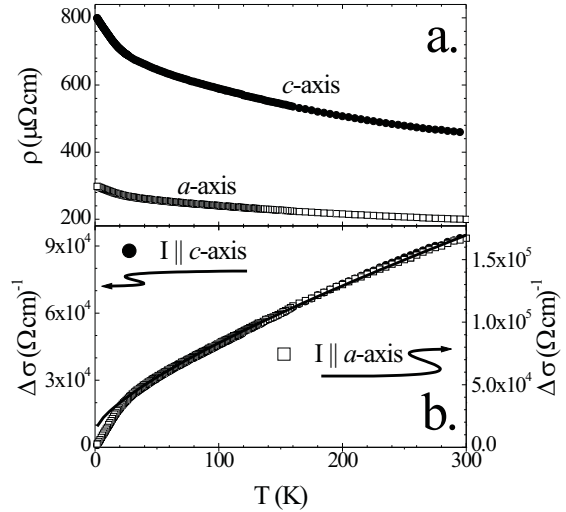}
\end{center}
\vspace{-0.6cm}\caption{(a) The resistivity $\rho$ and (b) the reduced conductivity $\Delta \sigma$ as function of temperature $T$ of single crystalline URh$_2$Ge$_2$, measured along the $a$ and $c$ axes of the tetragonal lattice. The solid line in (b) represents the result of a fit to the data; for details see text.}
\label{fig:fig1}
\end{figure}

For our experiments we have investigated an as--grown single crystalline specimen URh$_2$Ge$_2$ previously used in an XAFS/X--ray study~\cite{booth}. Other pieces of this single crystal, with slightly different absolute $\rho$ values, have been studied in Refs.~\cite{sullow1,sullow2}. The spin glass ground state below $T_f$\,$\simeq$\,9\,K has been established via susceptibility measurements. In Fig.~\ref{fig:fig1}(a) we plot the resistivity $\rho$, measured along the $a$ and $c$ axes. We observe the archetypical behavior of moderately disordered U heavy fermion compounds~\cite{sullow1}. The crystalline anisotropy is reflected in anisotropy of $\rho$ along $a$ and $c$ axes.

In Fig.~\ref{fig:fig1}(b) we plot the reduced conductivity $\Delta \sigma$ = $\sigma$\,--\,$\sigma_0$ as a function of $T$~\cite{sullow1}. After multiplying the $a$ axis data with a constant factor 1.77, $\Delta \sigma$ vs. $T$ for $a$ and $c$ axes superimpose over the full temperature range 1.5\,-\,300\,K. This proves that the electronic transport along the two directions is governed by the same mechanisms with a generic $T$ dependence.

The temperature dependence of $\Delta \sigma$ closely resembles the behavior of 3D amorphous metals. We quantify the resemblance by describing our data in terms of the corresponding localization theory~\cite{lee}, which successfully accounts for the conductivity of paramagnetic metallic glasses, amorphous ferromagnets or icosahedral U spin glasses~\cite{howson,das}. For weak electronic correlations (in URh$_2$Ge$_2$ at sufficiently high temperatures) the $T$ dependence of $\sigma$ is attributed to the superposition of incipient localization, destroyed by inelastic scattering with phonons and electrons, and electronic interaction effects. It is given by~\cite{howson}
\begin{equation}
\Delta \sigma (T) = \frac{e^2}{2\pi^2\hbar} \left( 3 \sqrt{b + c^2
T^2} - c T - 3 \sqrt{b} + d \sqrt{T} \right), \label{eq:eq1}
\end{equation}
with fit parameters $b=1/D \tau_{so}$, $c=\sqrt{1/4D\beta}$, $\beta=\tau_i T^2$~\cite{note1}, $d=0.7367\sqrt{ k/ D \hbar}$ (diffusion coefficient $D$, spin--orbit $(\tau_{so})$ and inelastic $(\tau_i)$ scattering times). Above 30\,K the $T$ dependence of $\Delta \sigma$ is well described by Eq.~\ref{eq:eq1}, thus validating our statement on the close resemblance to the behavior of amorphous metals. In Fig.~\ref{fig:fig1}(b) we include the result of a fit to the data as a solid line, using parameters~\cite{note2} along $a$/$c$ axis of $D = 0.54(20)/1.7(5) \times 10^{-6} m^2/s$; $\tau_{so} = 82(68)/82(56) \times 10^{-12} s$; $\beta = 50(29)/38(15) \times 10^{-10} s K^{2}$. The values of the diffusion coefficient $D$ are smaller by about an order of magnitude than those of common metallic glasses~\cite{howson}. With the Einstein relation $\sigma_0$\,=\,$e^2$\,$D$\,$N(E_f)$, it reflects the much larger density of states $N(E_f)$ in URh$_2$Ge$_2$~\cite{sullow1}. Conversely, the inelastic scattering times $\tau_i$ are larger by an order of magnitude in URh$_2$Ge$_2$, implying that the inelastic diffusion lengths $L_i$ are of the same order of magnitude as in metallic glasses ($L_i$\,=\,$\sqrt{D \tau_i}$\,$\sim$\,$O(10^{-7})m$ at low $T$)~\cite{howson}.

\begin{figure}[!ht]
\begin{center}
\includegraphics[width=1\linewidth]{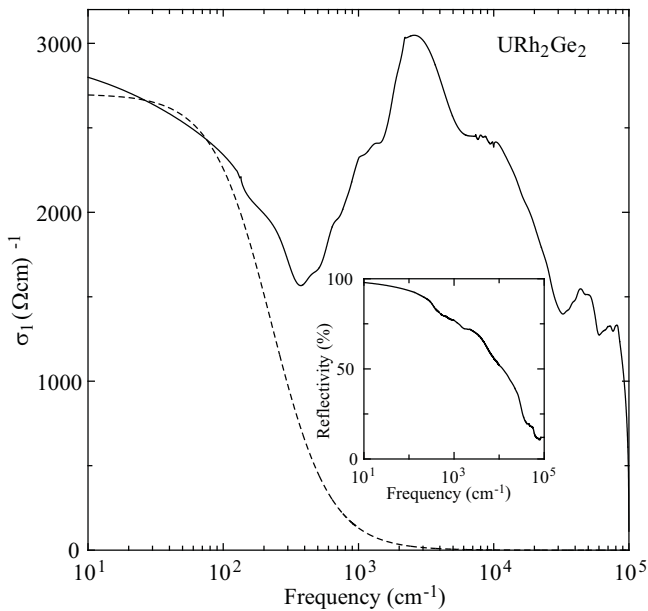}
\end{center}
\vspace{-0.6cm}\caption{The real part $\sigma_1$ of the $a$ axis optical conductivity of URh$_2$Ge$_2$ at 10\,K. The dashed line represents the result of a fit of the Drude component to the spectrum. The inset depicts the corresponding reflectivity spectrum.} \label{fig:fig2}
\end{figure}

In applying Eq.~\ref{eq:eq1} to URh$_2$Ge$_2$ we assume a metallic carrier density. To verify this assumption we have performed optical conductivity and Hall effect measurements. In Fig.~\ref{fig:fig2} we plot the result of the optical conductivity study, with the real part $\sigma_1(\omega)$ along the crystallographic $a$ axis obtained via Kramers--Kronig analysis of the reflectivity (inset of Fig.~\ref{fig:fig2}). Within experimental resolution, there is no temperature dependence of $\sigma_1$, reflecting the weak overall temperature dependence of $\rho(a)$ (Fig.~\ref{fig:fig1}(a)). The extrapolation of $\sigma_1$ to zero frequency yields a value of about 3000\,$(\Omega cm)^{-1}$, in good agreement with the value extracted from $\rho^{-1}_0(a)$.

The overall behavior of $\sigma_1(\omega)$ qualitatively and quantitatively resembles that of other heavy fermion metals~\cite{cao}, with a Drude--like conductivity at low frequencies and a maximum in $\sigma_1$ in the mid--infrared regime from optical inter--band excitations. This observation verifies the metallic ground state of URh$_2$Ge$_2$, and disproves semiconductor or Kondo insulator scenarios to account for the electronic transport properties of URh$_2$Ge$_2$. In consequence, the negative TCR must be the result of the crystallographic disorder.

\begin{figure}[!ht]
\begin{center}
\includegraphics[width=1\linewidth]{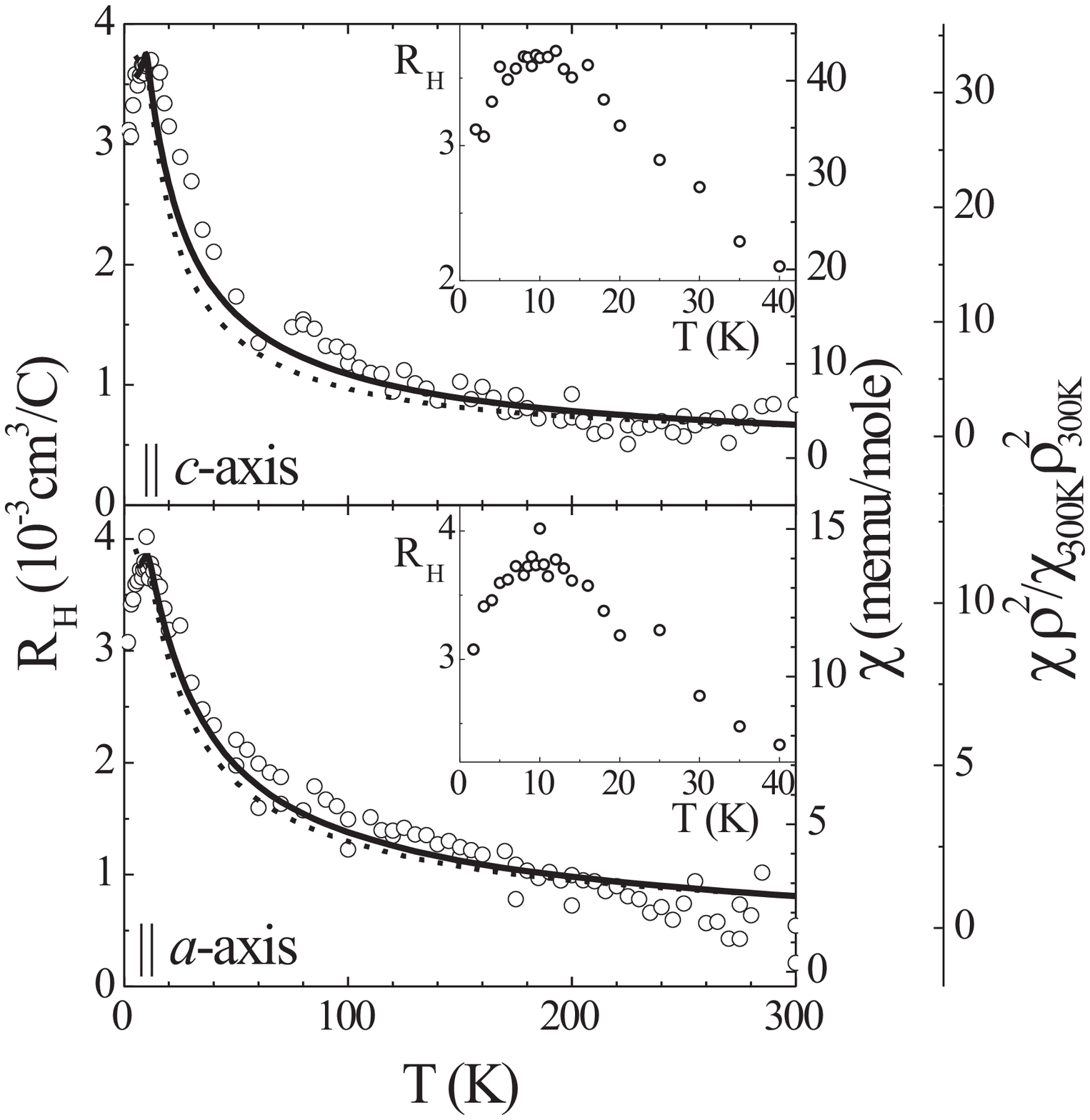}
\end{center}
\vspace{-0.6cm}\caption{The Hall constant $R_H$ of URh$_2$Ge$_2$ along $a$ and $c$ axes. The insets depict the low temperature regime, with the spin glass freezing transition causing the cusp like anomaly. The magnetic susceptibilities $\chi$ and the products $\chi \rho^2$ are displayed as solid and dashed lines, respectively; for details see text.} \label{fig:fig3}
\end{figure}

To quantitatively characterize the metallic ground state, we have measured the Hall constant $R_H$ along $a$ and $c$ axes of URh$_2$Ge$_2$ (Fig.~\ref{fig:fig3}). Here, the Hall constant has been derived from data taken below 1\,T. In this field range, both the Hall constant as the magnetization are linear in field. For both directions $R_H$ is dominated by anomalous Hall contributions. This is illustrated in Fig.~\ref{fig:fig3} by including the magnetic susceptibility $\chi=M/H$ (measured in a field of 0.05\,T). Down to lowest temperatures the $T$ dependence of $R_H$ is essentially that of $\chi$, {\it i.e.}, $R_H = R_0 + \chi R_S$ ($R_0$: ordinary Hall contribution; $R_S$: anomalous Hall coefficient). Assuming a spherical Fermi surface in a one band model, from these data we derive a carrier density $n$ of $\sim$\,3 carriers per unit cell for both axes.

Starting from the periodic Kondo lattice Anderson model, Kontani and Yamada~\cite{kontani} predicted a $T$ dependence of the anomalous Hall contribution in heavy fermions $\propto \chi$ above the coherence temperature $T_{coh}$. In URh$_2$Ge$_2$ coherence is suppressed with the crystallographic disorder, and thus $T_{coh}$\,=\,0. Accordingly, the prediction of Ref.~\cite{kontani} for $T$\,$>$\,$T_{coh}$ would describe our experimental observation.

Alternatively, in disordered media the side jump effect contributes to the anomalous Hall contribution such that $R_H$\,=\,$R_0$\,+\,$\chi$\,$\rho^2$\,$R_S$~\cite{berger}, with $\rho$ as the total electrical resistivity. In Fig.~\ref{fig:fig3} we include the normalized $T$ dependence of $\chi$\,$\rho^2$ (dashed lines). It also reproduces the overall behavior of $R_H$ and yields a carrier density of 1\,--\,2 carriers per unit cell.

Conversely, $R_H(T)$ of URh$_2$Ge$_2$ clearly is at variance with the Skew scattering prediction of Fert and Levy~\cite{fert}, $R_H$\,=\,$R_0$\,+\,$\chi$\,$\rho_{mag}$\,$R_S$ ($\rho_{mag}$: magnetic resistive contribution). Within this model, $\rho_{mag}$ approaches zero for $T \rightarrow 0$, resulting in a drastic decrease of $R_H$ at low temperatures and which is not observed in URh$_2$Ge$_2$.

Integrating the optical conductivity (Fig.~\ref{fig:fig2}) with respect to the effective metallic (Drude) component ({\it i.e.}, from 0 up to about 1000\,$cm^{-1}$) we obtain a plasma frequency of about 7000\,$cm^{-1}$. Assuming a band mass $m_b$\,=\,$m_e$ we estimate a free charge carriers concentration of about 5\,$\times$\,10$^{20}$\,$cm^{-3}$. The corresponding estimated $R_H$ constant is about a factor of 10 larger than measured. Such a discrepancy can be easily accounted for by a too small band mass as well as by an underestimation of the effective free charge carriers spectral weight. Extending the integration of $\sigma_1(\omega)$ up to about 1\,eV would lead to a perfect agreement, within the simple assumption $m_b$\,=\,$m_e$, with the measured Hall constant (the dashed line in Fig.~\ref{fig:fig2} represents the corresponding fit of the Drude component to the conductivity of URh$_2$Ge$_2$).

To further characterize the electronic transport in URh$_2$Ge$_2$ we have performed longitudinal and transversal magnetotransport experiments between 2 and 300\,K in fields up to 5\,T. As a representative, in Fig.~\ref{fig:fig3} we depict the longitudinal magnetoresistivity (MR) for the magnetic field $B$\,$\|$\,$c$, $(\rho(B)-\rho(B=0))/\rho(B=0)=\Delta \rho / \rho$, of URh$_2$Ge$_2$. For the other experimental geometries the size of the MR is smaller by one order of magnitude than for the case depicted in Fig.~\ref{fig:fig3}, but otherwise very similar in behavior~\cite{igor}. At all temperatures, for fields up to 5\,T the MR is small ($<$\,0.7\,\%) and follows essentially a $B^2$--dependence. While at low $T$, the magnetoresistivity is negative, surprisingly, it changes sign as temperature increases to above $\sim$\,100\,K.

\begin{figure}[!ht]
\begin{center}
\includegraphics[width=1\linewidth]{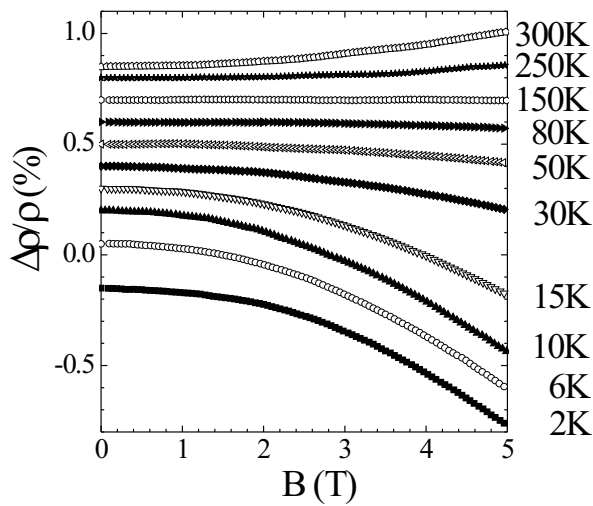}
\end{center}
\vspace{-0.6cm}\caption{The longitudinal magnetoresistivity of single crystalline URh$_2$Ge$_2$ between 2 and 300K, measured for $B \| c$ axis. Data are offset for clarity.} \label{fig:fig4}
\end{figure}

We have previously demonstrated that the negative MR at low $T$ stems from the reduction of spin disorder scattering~\cite{sullow2}. In contrast, a fundamentally different process is required to account for the positive MR at high $T$. For metallic glasses, in the limit $g$\,$\mu_B$\,$B$\,/\,$k_B$\,$T$\,$\ll$\,1 both spin--orbit and interaction effects yield a positive $\Delta \rho / \rho$\,$\propto$\,$B^2$~\cite{lee,dai}, just as observed for URh$_2$Ge$_2$ above $\sim$\,100\,K. While a full quantitative analysis fails because of parameter interdependency, with the above values for $D$ and $\beta$ the magnitude of the MR at $T$\,$\gtrsim$\,100\,K is reproduced using reasonable values for the interaction parameter $\gamma$\,$F_{\sigma}$ between 0 and 3~\cite{dai}. Therefore, we ascribe the positive MR at high $T$ to incipient localization effects and electronic interactions in the spirit of Eq.~\ref{eq:eq1}~\cite{lee,howson}.

To summarize, with our study on URh$_2$Ge$_2$ we have for the first time fully characterized the electronic transport properties of a moderately disordered heavy fermion compound. Our quantitative analysis demonstrates that the electronic transport in our single crystalline strongly correlated electron system with moderate, non--magnetic site disorder closely resembles the behavior of metallic glasses. Thus, we find an amorphous behavior in an essentially crystalline material, reflecting the enhancement of disorder effects with electronic correlations.

We acknowledge fruitful discussions with A. Castro--Neto, E. Miranda and W. Brenig. This work has been supported by the DFG under contract no. SU229/1-3.


\begin{references}
\bibitem{wysokinski} K.I. Wysokinski, Phys. Rev. B {\bf 60}, 16376 (1999).
\bibitem{maple} M.C. de Andrade {\it et al.}, Phys. Rev. Lett. {\bf 81}, 5620 (1998).
\bibitem{capogna} L. Capogna {\it et al.}, Phys. Rev. Lett.{\bf 88}, 076602 (2002); Z.Q. Mao {\it et al.}, Phys. Rev. Lett. {\bf 90}, 186601 (2003); Y.B. Kim, A.J. Millis, Phys. Rev. B {\bf 67}, 085102 (2003).
\bibitem{millis} A.J. Millis, Solid State Commun. {\bf 126}, 3 (2003).
\bibitem{belitz} D. Belitz and T.R. Kirkpatrick, Rev. Mod. Phys. {\bf 66}, 261 (1994); M. Imada, A. Fujimori, and Y. Tokura, Rev. Mod. Phys. {\bf 70}, 1039 (1998).
\bibitem{dobro} V. Dobrosavljevic, T. R. Kirkpatrick, and G. Kotliar, Phys. Rev. Lett. {\bf 69}, 1113 (1992); V. Dobrosavljevic and G. Kotliar, Phys. Rev. Lett. {\bf 78}, 3943 (1997).
\bibitem{stewart} G.R. Stewart, Rev. Mod. Phys. {\bf 73}, 797 (2001).
\bibitem{rosch} A. Rosch, Phys. Rev. Lett. {\bf 82}, 4280 (1999).
\bibitem{miranda1} E. Miranda, V. Dobrosavljevic, and G. Kotliar, Phys. Rev. Lett. {\bf 78}, 290 (1997); A.H. Castro Neto, G. Castilla, and B.A. Jones, Phys. Rev. Lett. {\bf 81}, 3531 (1998).
\bibitem{andraka} B. Andraka and G.R. Stewart, Phys. Rev. B {\bf 47}, 3208 (1993); R.P. Dickey {\it et al.}, Phys. Rev. B {\bf 62}, 3979 (2000).
\bibitem{shlyk} L. Shlyk and J. Stepien-Damm, J. Magn. Magn. Mat. {\bf 195}, 37 (1999).
\bibitem{sullow1} S. S\"{u}llow {\it et al.}, Phys. Rev. B {\bf 61}, 8878 (2000); J. Magn. Magn. Mat. {\bf 226--230}, 35 (2001); {\it ibid,} {\bf 272--276}, 954 (2004).
\bibitem{li} D.X. Li {\it et al.}, J. Phys. Soc. Jpn. {\bf 71}, 418 (2002).
\bibitem{miranda2} M.C.O. Aguiar, E. Miranda, and V. Dobrosavljevi\'c, Phys. Rev. B {\bf 68}, 125104 (2003).
\bibitem{lee} P.A. Lee and T.V. Ramakrishnan, Rev. Mod. Phys. {\bf 57}, 287 (1985); B.L. Altshuler and A.G. Aronov, in {\it Electron--Electron Interaction in Disordered Systems}, ed. by A.L. Efros and M. Pollak (Elsevier, New York, 1985), p. 1.
\bibitem{sullow2} S. S\"{u}llow {\it et al.}, Phys. Rev. Lett. {\bf 78}, 354 (1997).
\bibitem{booth} C.H. Booth {\it et al.}, J. Magn. Magn. Mat., {\bf 272--276}, 941 (2004); C.H. Booth, R. Feyerherm, S. S\"{u}llow, in preparation.
\bibitem{howson} B.J. Hickey, D. Greig, and M.A. Howson, J. Phys. F: Met. Phys. {\bf 16} (1986) L13; H.H. Boghosian and M.A. Howson, Phys. Rev. B {\bf 41}, 7397 (1990).
\bibitem{das} A. Das and A.K. Majumdar, Phys. Rev. B {\bf 43}, 6042 (1991); K.M. Wong and S.J. Poon, Phys. Rev. B {\bf 34}, 7371 (1986).
\bibitem{note1} Since $\tau_i$ is controlled by phonon scattering, {\it i.e.}, $\tau_i$\,$\propto$\,$T^{-2}$, the coefficient $\beta$ is independent of temperature.
\bibitem{note2} The large error bars of the fit parameters reflect interdependencies, prohibiting a more detailed analysis including the interaction parameter $\tilde{F}$, which was fixed at 0.
\bibitem{cao} N. Cao {\it et al.}, Phys. Rev. B {\bf 53}, 2601 (1996); L. Degiorgi, Rev. Mod. Phys. 71, 687 (1999); L. Degiorgi, F.B.B. Anders, G. Gr\"{u}ner, Eur. Phys. J. B {\bf 19}, 167 (2001); M. Dressel {\it et al.}, Phys. Rev. Lett. {\bf 88}, 186404 (2002).
\bibitem{kontani} H. Kontani and K. Yamada, J. Phys. Soc. Jpn. {\bf 63}, 2627 (1994).
\bibitem{berger} L. Berger, Phys. Rev. B {\bf 2}, 4559 (1970); F.M. Mayeya and M.A. Howson, Phys. Rev. B {\bf 49}, 3167 (1994).
\bibitem{fert} A. Fert and P.M. Levy, Phys. Rev. B {\bf 36}, 1907 (1987).
\bibitem{igor} I. Maksimov, Ph.D. thesis, TU Braunschweig (2003), unpublished.
\bibitem{dai} P. Dai, Y. Zhang, and M.P. Sarachik, Phys. Rev. B {\bf 46}, 6724 (1992).
\end{references}
\end{document}